\documentclass[preprint,12pt]{elsarticle}

\usepackage{amssymb}
\usepackage{graphicx}

\def\a  {\alpha}       \def\b  {\beta}         \def\g  {\gamma}
       \def\d  {\delta}        
        
\def\l  {\lambda}

 \newcommand{\caln}{\mbox{${\cal N}$}}

\def\IR{{\hbox{{\rm I}\kern-.2em\hbox{\rm R}}}}
\def\IB{{\hbox{{\rm I}\kern-.2em\hbox{\rm B}}}}
\def\IN{{\hbox{{\rm I}\kern-.2em\hbox{\rm N}}}}
\def\IC{\,\,{\hbox{{\rm I}\kern-.59em\hbox{\bf C}}}}
\def\IZ{{\hbox{{\rm Z}\kern-.4em\hbox{\rm Z}}}}
\def\IP{{\hbox{{\rm I}\kern-.2em\hbox{\rm P}}}}
\def\IH{{\hbox{{\rm I}\kern-.4em\hbox{\rm H}}}}
\def\ID{{\hbox{{\rm I}\kern-.2em\hbox{\rm D}}}}

\def\be{\begin{equation}}
\def\ee{\end{equation}}
\def\ba{\begin{eqnarray}}
\def\ea{\end{eqnarray}}

\newcommand{\inv}[1]{\frac{1}{#1}}

\def\ra{\rightarrow}

\def\dell{\partial}

\def\Tr{{\rm tr}\,}

\def\det{{\rm det}}

\def\nn{\nonumber}
\def\ea{{\it et al}. }

\newcommand{\ie}{{\it i.e.}}

\newcommand{\gym}{g_{Y\!M}}

\def\De{\textrm{D8}}
\def\DeB{\overline{\textrm{D8}}}

\newcommand{\Ukk}{U_{\rm KK}}
\newcommand{\Mkk}{M_{\rm KK}}

\begin{document}

\begin{frontmatter}



\title{On the Baryonic Density and Susceptibilities in a Holographic Model of QCD}

\author[kk]{Keun-young Kim}
\author[jl]{Jinfeng Liao}

\address[kk]{Department of Physics \& Astronomy, SUNY Stony Brook, NY 11794, USA.}
\address[jl]{Nuclear Science Division, Lawrence Berkeley National Laboratory, \\
MS70R0319, 1 Cyclotron Road, Berkeley, CA 94720, USA.}

\begin{abstract}
In this paper, we calculate analytically the baryonic density and
susceptibilities, which are sensitive probes to the fermionic
degrees of freedom, in a holographic model of QCD both in its hot
QGP phase and in its cold dense phase. Interesting patterns due to
strong coupling dynamics will be shown and valuable lessons for
QCD will be discussed.
\end{abstract}

\begin{keyword}
Holographic QCD \sep AdS/CFT \sep Baryonic Susceptibilities


\end{keyword}

\end{frontmatter}

\newpage


\setcounter{page}1

\section{Introduction}

Historically string theory was developed during the attempt to
build dual models for the strong interaction describing various
hadrons and their scattering amplitudes \cite{Schwarz:2007yc}.
After the advent of Quantum Chromodynamics (QCD) as the
fundamental gauge theory of strong interaction, string theory
departed into a quite different route with many fascinating
discoveries. More recently, a new twist has been added to the
story between string theory and QCD: the AdS/CFT correspondence
\cite{Maldacena:1997re,Gubser:1998bc,Witten:1998qj}, or more
generally the gauge/gravity duality, has provided a way to study
the strong coupling dynamics of non-Abelian gauge theories (like
QCD) via its string theory dual in the much more tractable
supergravity regime. A lot of interesting results have been
achieved along this direction with applications in various
aspects, for reviews see e.g.
\cite{Aharony:1999ti,Peeters:2007ab,Mateos:2007ay,Erdmenger:2007cm,Alday:2008yw,
Gubser:2009md,Gubser:2009sn,Rangamani:2009xk}.

While a specific gravity dual for real QCD hasn't been found yet,
many important insights into the nonperturbative aspects of QCD
have been obtained via gauge/gravity duality. Let's just mention
two remarkable examples of such kind. Study of the so-called
quark-gluon plasma (QGP), the deconfined phase of QCD at
temperature higher than the deconfinement transition $T>T_c\approx
170-200MeV$ and low baryonic density, has been very important but
proved to be rather difficult. Two powerful tools for such study
include the lattice QCD and the heavy ion collisions experiments
(e.g. at RHIC facility): interestingly, the former approach found
that the the QGP thermodynamics (e.g. the energy density) deviates
from Stefan-Boltzmann limit constantly by about 20\% in the range
$1.5-4T_c$ \cite{Karsch:2001cy}, while the latter approach found
that the QGP produced at RHIC (corresponding to $1-2T_c$) has an
extremely small shear viscosity to entropy density ratio $\eta/s$
of the order $0.1$ \cite{sQGP_review}. Both discoveries challenged
naive expectation of a weakly coupled QGP right above $T_c$ and
led to the paradigm shift to a strongly coupled QGP (sQGP)
\cite{sQGP_review,Shuryak_review,Kharzeev:2009zb}. It turns out
that both results can be much better understood in light of the
strong coupling results from gauge/gravity duality: the
thermodynamics of CFT plasma at infinitely strong coupling has
been calculated via its dual black hole to be exactly
$\frac{3}{4}$ of the Stefan-Boltzmann limit \cite{Gubser:1996de},
while a large class of strong coupling gauge theories with gravity
dual have been shown to have their
$\frac{\eta}{s}=\frac{1}{4\pi}\approx 0.08$
\cite{Policastro:2001yc}. These successes have inspired a lot of
activity to extract useful information for QCD and QGP via
gauge/gravity duality, see e.g. reviews in
\cite{Gubser:2009md,Gubser:2009sn,Shuryak_review}.

In this paper, we focus on the fermionic sector i.e. the degrees
of freedom carrying baryonic charges in QCD thermodynamics, and
use the gauge/gravity duality to qualitatively obtain some of
their nonperturbative features. There are very rich dynamics
associated with the quarks in QCD, for example the spontaneous
chiral symmetry breaking in QCD vacuum and its restoration at high
temperature/density. There are also interesting phase structures
in the low T and high baryonic density region of QCD phase diagram
where the fermionic sector becomes dominant. Color
superconductivity \cite{Rapp:1997zu} was found to occur at very
high density with the phase presumably a color-flavor-locking one
\cite{Alford:1998mk}. At moderately high density there are
interesting phenomena like e.g. the interplay between the color
superconductivity and chiral restoration and the pairing with
mismatched Fermi surfaces \cite{Berges:1998rc}\cite{Liao:2003qw}.
More recently based on wisdom from large $N_c$ argument, it has
been proposed that there could be a new phase in the cold dense
region, named ``quarkyonic'' phase
\cite{McLerran:2007qj}\cite{McLerran:2008ua}, which is confined
and yet has thermodynamics scaling as $N_c$ like a system made of
quarks. There are also many recent results on the baryonic density
and susceptibilities from lattice QCD which show nonperturbative
patterns
\cite{Allton:2005gk,Cheng:2008zh,Gavai:2005sd,Bernard:2007nm,Hietanen:2008xb}.

It is thus of great interest to study these aspects of QCD with
the handy tool of gauge/gravity duality. To this end, we use the
Sakai-Sugimoto(SS) model \cite{Sakai}, which have reproduced (at
lease qualitatively) an impressive number of QCD results, for
example: the baryon property, form factor~\cite{Baryon}, and
the nuclear force~\cite{NuclearForce}, etc. The SS model has been
studied at finite temperature~\cite{FiniteT} and at finite baryon
density in~\cite{FiniteD,Bergman0,Kim:Dense}. We will particularly
calculate the baryonic density and susceptibilities, which are
sensitive probes to the fermionic degrees of freedom, from the
Sakai-Sugimoto model both in the hot QGP phase and in the cold
dense phase with the motivation to understand their patterns under
strong coupling and find valuable lessons for QCD.

The paper is organized as follows. We give in Section.II a brief
review of baryonic density and susceptibilities in QCD and in
Section.III a brief review of pertinent Sakai-Sugimoto results
obtained before. The baryonic density and susceptibilities in
Sakai-Sugimoto model will be analytically calculated for the hot
QGP phase in Section.IV and for the cold dense phase in Section.V.
Finally in Section.VI we summarize the results and discuss
relevant lessons for QCD.

\section{Brief Review of Baryonic Density and Susceptibilities in QCD}

\subsection{Definition and Examples }
We start with a Taylor expansion of pressure $P(T,\mu)$ with
respect to chemical potential $\mu$ (with the convention that
quark carries unit baryonic charge) at fixed $T$:
\begin{equation}
P(T,\mu)=T^4 \sum_{n=0}^{\infty} \frac{d_n(T)}{n!}
\left(\frac{\mu}{T}\right)^n
\end{equation}
with the (dimensionless) baryonic susceptibilities $d_n(T)$
defined as
\begin{equation} \label{eqn_susceptibilities_def}
d_n(T) \equiv {\partial^n (P/T^4)\over \partial
  (\mu/T)^n}{\bigg |}_{\mu=0}
\end{equation}
Note for the above $d_n$ the odd-$n$ ones vanish by symmetry.
Furthermore we see all non-zero $d_n$ except $n=0$ represent
certain contribution from the baryonic degrees of freedom, and
importantly all non-baryonic degrees of freedom (e.g. the gluonic
sector in QCD) do {\em not} directly contribute to them. So these
derivatives probe the properties of the effective fermions in the
system directly and sensitively. This can be seen also from the
e.g. the baryonic density as given by
\begin{equation}
n_B(T,\mu)= T^3 {\partial (P/T^4)\over \partial
  (\mu/T)} = T^3 \sum_{n=2}^{\infty} \frac{d_n(T)}{(n-1)!}
\left(\frac{\mu}{T}\right)^{n-1}
\end{equation}

To give an idea and to provide a benchmark of the baryonic density
and susceptibilities, we explicitly evaluate these for a free gas
of particles with mass $M$ and baryonic charge $B$. The density is
given by
\begin{equation}
n_B^{free}(T,\mu)=B N_i \frac{T^3}{2\pi^2} \int_0^\infty dx
\left\{ \frac{x^2}{e^{\sqrt{(\beta M)^2 + x^2}-(\beta B \mu)}+1} -
\frac{x^2}{e^{\sqrt{(\beta M)^2 + x^2}+(\beta B \mu)}+1} \right\}
\end{equation}
with $\beta=1/T$ and $N_i$ denoting the number of internal degrees
of freedom (e.g. color,flavor,spin, etc). The susceptibilities
$d_n$ can be obtained by subsequent differentiation with respect
to $\mu$. We are particularly interested in two limiting cases:
the non-relativistic(NR) limit with $\beta M \to \infty$ and the
ultra-relativistic(UR) limit with $\beta M \to 0$. In these limits
we can obtain concrete results below.\\
\textbf{(i) NR limit}: the baryonic density and susceptibilities
in this limit are:
\begin{equation}
n_B^{free}{\bigg |} _{NR} = B  N_i \left(MT \over 2\pi
\right)^{3\over 2} e^{- {M\over T}} \left[ e^{B\mu \over T} -
e^{-{B\mu \over T}} \right]
\end{equation}
\begin{equation} \label{eqn_NR_d}
d_{n}^{free}{\bigg |} _{NR} = N_i \left(M \over 2\pi T
\right)^{3\over 2} e^{- {M\over T}} \times 2 B^{n} \equiv
{\mathcal F}\left[{M\over T}\right] B^{n}
\end{equation}
We observe a few important points at this limit: (\textbf{a}) all
susceptibilities $d_{n}$ are {\em positive} and have the {\em same
dependence on $T$} up to a constant coefficient ; (\textbf{b}) the
ratio between successive susceptibilities is directly related to
the baryonic charge carried by the degree of freedom, i.e.
$d_{n+2}/d_n=B^2$, independent of $M,T$ and $n$; (\textbf{c}) for
multi-component non-interacting gas of species $M_i,B_i$, all
$d_{n}$ are simply a sum over species with the same formulae above
and they remain all {\em positive}, but one then expect the ratios
$d_{n+2}/d_n=<B_i^2>$ to be an abundance-averaged baryonic charge
which now depends on $B_i$ and $M_i,T,n$ as well. \\
\textbf{(ii) UR limit}: the baryonic density and susceptibilities
in this limit are:
\begin{equation}
n_B^{free}{\bigg |} _{UR} = N_i  \frac{T^3}{6\pi^2} \left[
B^4\left({\mu\over T} \right)^3 + \pi^2 B^2 \left({\mu\over
T}\right) \right]
\end{equation}
\begin{equation} \label{eqn_UR_d}
d_2^{free}{\bigg |} _{UR} = N_i \frac{B^2}{6} \quad , \quad
d_4^{free}{\bigg |} _{UR} = N_i \frac{B^4}{\pi^2} \quad , \quad
d_{n>4}^{free}{\bigg |} _{UR} = 0
\end{equation}
Again for multi-component non-interacting gas of species
$M_i,B_i$, the $d_{n}$ are simply a sum over species with the same
formulae above. For such a Stefan-Boltzmann gas of quarks (with
$B_q=1$ convention) with spin $N_s$, flavor $N_f$ and color $N_c$,
the susceptibilities per degrees of freedom (D.o.F) are simply
$d_2^{SB}/(N_s N_f N_c)=\frac{1}{6}$, $d_4^{SB}/(N_s N_f
N_c)=\frac{1}{\pi^2}$ with all higher ones vanishing.

In both limits of the free gas example above, we find all
non-vanishing susceptibilities to be positive and proportional to
$B^n$ which implies the contribution of fermions with large $B$
becomes larger and larger with increasing order $n$.

We end this part by emphasizing again that the density and
susceptibilities are direct and sensitive probes to the fermions
with baryonic charges in the system. In particular their deviation
from the free patterns encodes important information about the
dynamics and it is certainly of great interest to know the {\em
behavior of baryonic density and susceptibilities under strong
coupling}.

\subsection{The Susceptibilities in Lattice QCD}

While the susceptibilities contain very useful information about
the fermionic degrees of freedom, it is generally hard to be
calculated in a strongly coupled theory like QCD. Nevertheless
such susceptibilities of QCD can be studied in its lattice
formulation. In particular, calculating the $d_n$ (defined at {\em
zero} $\mu$) offers an important method (via Taylor expansion) to
explore the QCD phase diagram at finite density which was
traditionally unaccessible for lattice QCD due to the well-known
``sign problem''. These susceptibilities are also directly related
to baryonic charge fluctuations in the thermal QCD matter created
in heavy ion collisions which can be experimentally measured
\cite{Koch:2008ia}. The first lattice results for 2-flavor QCD
\cite{Allton:2005gk} with a relatively
 large pion mass by Karsch et al from a few years ago showed highly
 nontrivial patterns near $T_c$, with strong deviation from the free case in the
 region $T\sim 1-2T_c$. More recent 3-flavor results with a more realistic pion mass
 by the same group \cite{Cheng:2008zh} still preserved the nontrivial patterns
 close to $T_c$ though with deviation from the free case limited to be below $\sim 1.4Tc$.
 See also other lattice works in e.g. \cite{Gavai:2005sd}\cite{Bernard:2007nm}\cite{Hietanen:2008xb}
 which all show similar behavior of the susceptibilities.

\begin{figure*}
 \center{ \hskip 0in\includegraphics[width=4.2cm]{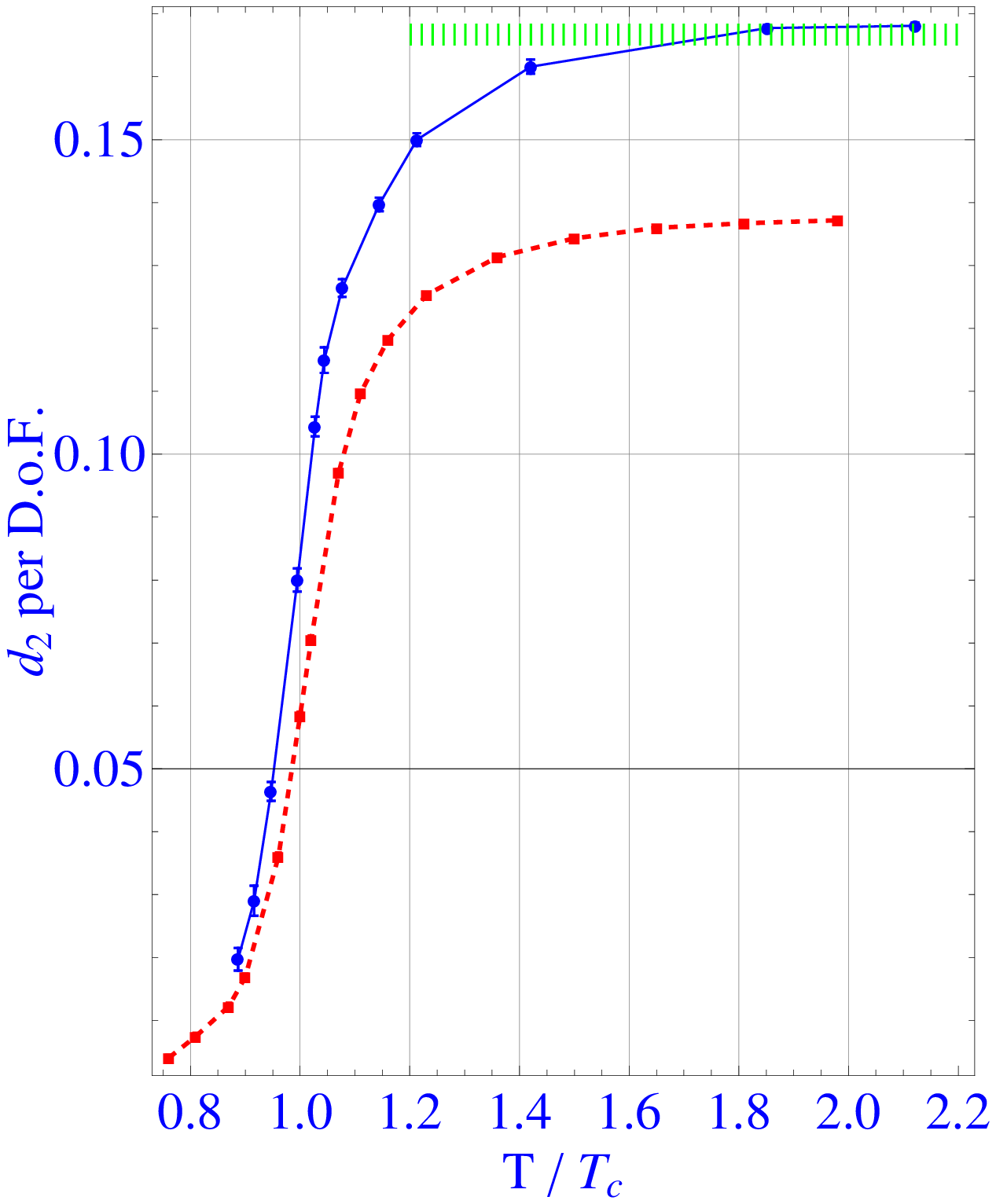}
    \hskip 0.05in\includegraphics[width=4.2cm]{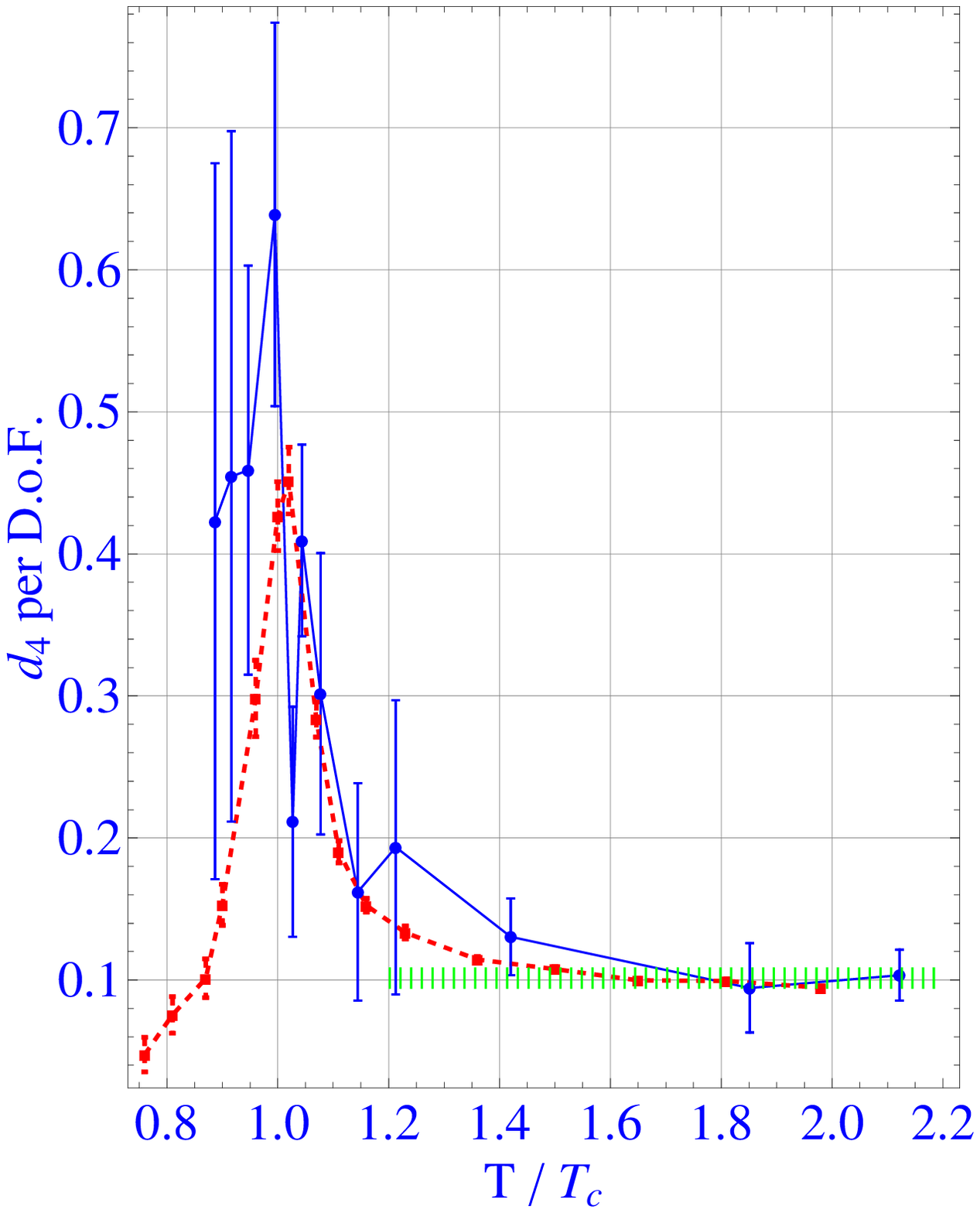}
      \hskip 0.05in\includegraphics[width=4.2cm]{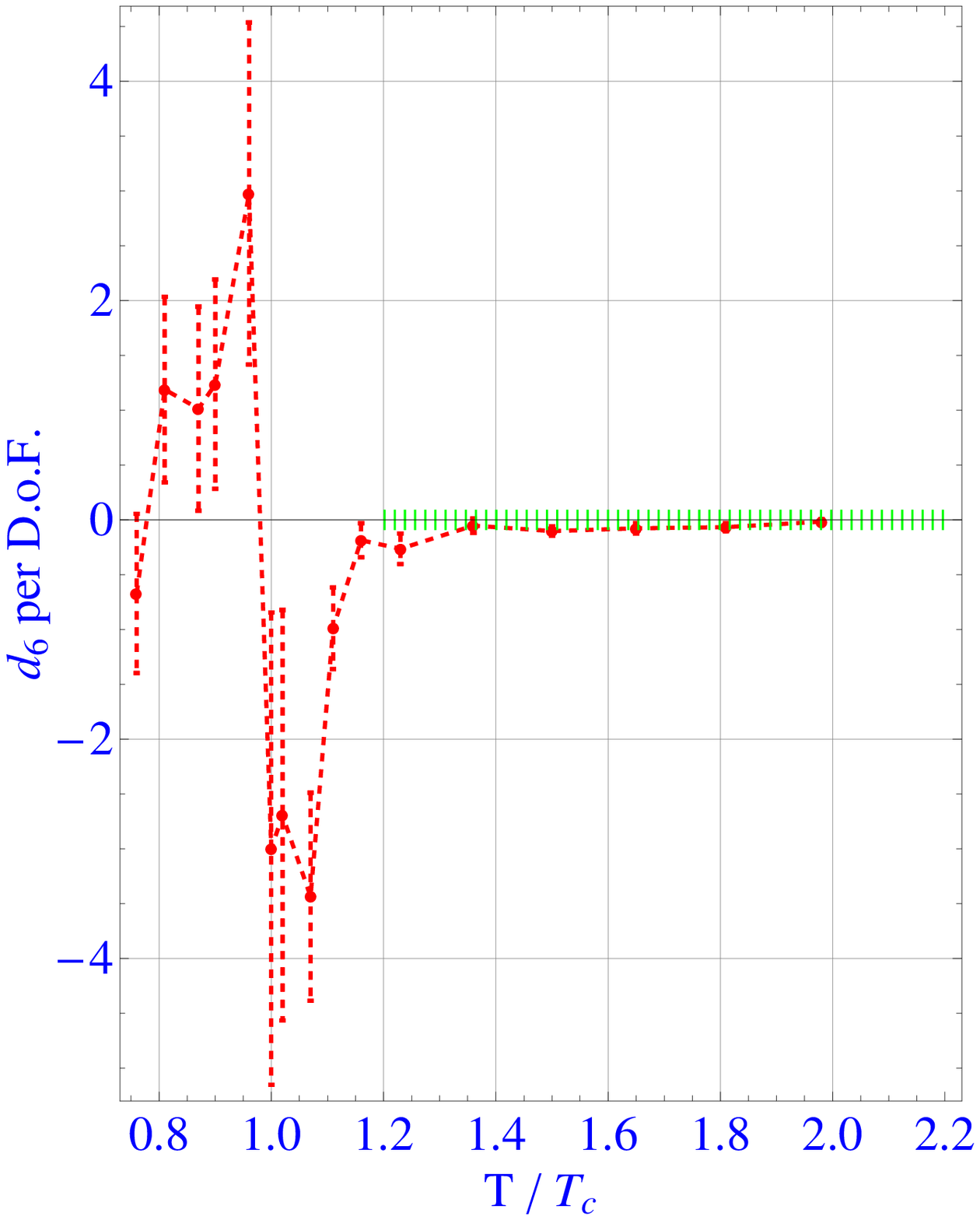}}
 \caption{\label{fig_lattice} (from left to right) the susceptibilities $d_2$,
 $d_4$, and $d_6$ per D.o.F. calculated from lattice QCD. The dashed
 red lines in all panels are earlier 2-flavor results in \protect\cite{Allton:2005gk} while
 the solid blue lines in $d_2$ and $d_4$ plots are recent 3-flavor results
 by the same group in \protect\cite{Cheng:2008zh}. The
 dotted green bands represent the Stefan-Boltzmann limit.}
 \end{figure*}

We now make a detailed discussion on these lattice results. Let's
start with the $T<T_c$ part, i.e. the confined hadronic phase: in
this phase we actually see from Fig.\ref{fig_lattice} the same
signs and similar T-dependence of $d_{2,4,6}$ (including
exponential-like growth close to $T_c$) hinting at the NR limit of
a free gas as in Eq.(\ref{eqn_NR_d}). And indeed a simple hadronic
resonance gas model including various baryons from the Particle
Data Book can nicely fit the lattice data for $T<T_c$, see e.g.
Fig.1 in \cite{Liao:2005pa}. In contrast, the data for $T>T_c$ do
{\em not} resemble any free gas (NR or UR) model at all. A few
nontrivial features are readily visible from
Fig.\ref{fig_lattice}, especially for $1-1.4T_c$: {\em $d_{2,4,6}$
have rather distinctive patterns, with $d_2$ positive and mildly
growing, $d_4$ positive and rapidly dropping, while $d_6$ negative
and dropping even more abruptly.} The susceptibilities results
clearly disfavor any weakly interacting quasiparticle models for
the deconfined quark-gluon plasma (QGP) phase right above $T_c$,
instead they indicate non-perturbative effect from strong
coupling. This observation turns out to be consistent with the
conclusion from other independent approaches that the QGP in the
same $1-2T_c$-region is rather strongly coupled, now called sQGP.
In particular the experimental study of the QGP phase by heavy ion
collisions at RHIC suggests that it behaves as a very good liquid
\cite{sQGP_review,Shuryak_review,Kharzeev:2009zb,Gubser:2009md}.
Lattice works on e.g. thermodynamics \cite{Cheng:2007jq} and
transport properties \cite{Meyer:2007ic}, heavy quark potentials
\cite{Kaczmarek_Zantow} and charmonium above $T_c$
\cite{charmonium} also point to a similar conclusion. A
microscopic explanation of such strong coupling results may rely
on understanding of the specific mechanism of how the confinement
occurs toward $T_c$ from above: it has been suggested under the
generic spirit of electric-magnetic duality that the QCD plasma
close to $T_c$ is actually a {\em magnetic} one, made of light and
abundant monopoles which become Bose-condensed at $T_c$ and
enforce confinement
\cite{Liao:2006ry,Chernodub,D'Alessandro:2007su,Ratti:2008jz}.

Returning to the non-trivial susceptibilities in $1-1.4T_c$, a
theoretical understanding is still lacking. One particular
suggestion involves possible bound states of quarks and gluons
even in the plasma phase due to the presence of still strong
coupling in $1-2T_c$ \cite{SZ_rethinking,SZ_bound,Liao:2005hj},
including both conventional colorless states (remnants of mesons
and baryons) and colored states like diquarks, q-g states and even
polymer-like chains. Those bound states carrying baryonic charges
can contribute substantially to the susceptibilities, as first
pointed out by Liao and Shuryak in \cite{Liao:2005pa}, and
especially become more and more important in higher orders. To see
this, one just notes that diquarks have baryonic charge $B=2$ and
baryons $B=3$, and that the susceptibilities go like $d_n \sim
B^n$ as evident from Eq.(\ref{eqn_NR_d},\ref{eqn_UR_d}): this
means even the density of the bound states may be small compared
to quarks but they still can dominate e.g. $d_4,d_6$ and even
higher order ones. It was shown in \cite{Liao:2005pa} that the
bound states contribution, mainly from baryons, could be dominant
for the peak and wiggle structures seen in the lattice data. This
was confirmed by an even better agreement with data in later
studies using the so-called PNJL model \cite{Ratti:2005jh} in
which due to Polyakov line suppression of colored states below and
around $T_c$, precisely the 3-quark colorless combinations (e.g.
baryons) dominate the nontrivial susceptibilities around $T_c$.
Despite these progresses, it is still unclear {\em how the strong
coupling dynamics affects the susceptibilities from
individual quarks} --- this is what we attempt to address in the present paper. \\

To conclude this section, we have seen that the baryonic density
and susceptibilities are sensitive probes to the degrees of
freedom carrying baryonic charges, and can exhibit rather
nontrivial structures in strong coupling regime as shown by
lattice QCD data. It is natural, then, to study these properties
for strongly coupled gauge theories in general and for useful
models of QCD in particular. To this end, the Sakai-Sugimoto model
\cite{Sakai} is a good choice: as a holographic model of QCD, it
is calculable in strong coupling regime by virtue of the
gauge/string duality and has been shown to have many realistic
features of QCD. In the rest of the paper, we will calculate the
baryonic density and susceptibilities in the Sakai-Sugimoto model
and discuss the lessons for QCD.

\section{Brief Review of Sakai-Sugimoto Model with Chemical Potential}

In this section we summarize the Sakai-Sugimoto (SS) model for
notation and completeness. For a thorough presentation we refer
\cite{Sakai} for zero temperature and \cite{FiniteT} for finite
temperature.

The SS model, in brief, is defined by the dynamics of $N_f$
$\De$-$\DeB$ branes in the background field (the metric, the dilaton,
and the Ramond-Ramond field) generated by $N_c$ D4-branes. In
order not to disturb the background we require $N_f \ll N_c$,
which is called the ``probe'' limit and corresponds to the quenched
approximation. The low energy dynamics of $\De$-$\DeB$ branes are
governed by the Dirac-Born-Infeld (DBI) action and the
Chern-Simons (CS) action:
\begin{eqnarray}
S_{\mathrm{DBI}}&=& -T_8 \int d^9 x \ e^{-\phi}\ \Tr
\sqrt{-\det(g_{MN}+2\pi\alpha' F_{MN})} \ , \label{DBI0}\\
S_{\mathrm{CS}}&=&\frac{1}{48\pi^3} \int C_3 \Tr F^3 \label{CS0} \
,
\end{eqnarray}
where $\phi$ and $C_3$ are the dilaton and the Ramond-Ramond
field. The metric generated by D4 brane is encoded in the induced
metric $g_{MN}$ on the $\De$-$\DeB$ branes. $F_{MN}=\partial_M A_N
-\partial_N A_M -i \left[ A_M , A_N \right]$ ($M,N =
0,1,\cdots,8$) are the field strength tensor of the $U(N_f)$ gauge
fields on the D8-branes. Note the $\Tr$ is taken over the
$N_f$-dimensional flavor space. $T_8 = \frac{1}{(2\pi)^8 l_s^9}$
is the tension of the D8-brane. In the following subsections we
will specify the ingredient fields of the DBI and CS action: the
induced metric, dilaton, RR field, and the gauge field.

\subsection{Ingredient fields}

The induced metric on the D8 branes from the D4 branes  background metric can be written as~\cite{Sakai,FiniteT}
\begin{eqnarray}
ds_{\mathrm{D8}}^2 = g_{00} (dX^E_0)^2 + g_{xx} (d\vec{X})^2
+ g_{UU} dU^2 + g_{SS}d\Omega_4^2  \ ,\label{D8metric}
\end{eqnarray}
where\footnote{In the appendix, the background metric and coordinate notation convention are shown.}
\begin{eqnarray}
 && g_{00} = \a\left(\frac{U}{R}\right)^{3/2} , \
  g_{xx} = \left(\frac{U}{R}\right)^{3/2} , \
  g_{UU} = \left(\frac{R}{U}\right)^{3/2} \g , \
  g_{SS} = \left(\frac{R}{U}\right)^{3/2}U^2 \  \nn \\
&&\ \a \ \ra\ 1 \  , \qquad  \g \ \ra\ \inv{f(U)} +
\left(\frac{\dell X_4}{\dell U}\right)^2 \left(\frac{U}{R}
\right)^3 f(U) \ ,
\quad\mathrm{(Confined\ phase)} \nn \\
&&\ \a \ra f(U) \  , \quad \g \ra \inv{f(U)} +
\left(\frac{\dell X_4}{\dell U}\right)^2 \left(\frac{U}{R}
\right)^3 \ . \qquad \quad \mathrm{(Deconfined\ phase)} \nn
\end{eqnarray}
The embedding information is  encoded only in $\gamma$ and thereby
$g_{UU}$. For definition of the warping factor $f(U)$ in both
confined and deconfined phases, see the appendix.
%Note that
%
%\begin{eqnarray}
%  g_{00}g_{UU} = \a\g =
% \left(\frac{U}{R} \right)^3 f(U) \left(\frac{\dell
%X_4}{\dell U}\right)^2 + \left\{\begin{array}{c} \inv{f(U)} \%\ 1
%\end{array} \right\}
%   \   \begin{array}{l} \mathrm{confined\ phase}
%   \\ \mathrm{deconfined\ phase} \end{array} \ .
%\label{gg}
%\end{eqnarray}
%

The dilaton and RR field are given as
\begin{eqnarray}
e^\phi= g_s \left(\frac{U}{R}\right)^{3/4}, ~~F_4\equiv
dC_3=\frac{2\pi N_c}{\Omega_4}\epsilon_4 \ ,
\end{eqnarray}
where $g_S$ is the string coupling constant, $\Omega_4=8\pi^2/3$ is the volume of the unit $S^4$ and
$\epsilon_4$ is the corresponding volume form.

For the gauge field we only consider the time component of the
$U(1)$ part of the $U(N_f)$ gauge field, which is normalized as
\begin{eqnarray}
  \frac{1}{\sqrt{2N_f}} \hat{A}_0(U) \ ,
\end{eqnarray}
where we also assumed the $\hat{A}_0$ is only a function of $U$. This
choice of the \emph{bulk} gauge field is a standard holographic
way to introduce the chemical potential in the \emph{boundary}
field theory. $\hat{A}_0(\infty)$ of the classical solution is
identified with the chemical potential.

\subsection{The DBI action}
the DBI action (\ref{DBI0}) reads
\begin{eqnarray}
  S_{\mathrm{DBI}} =
  N_s \cdot \Omega_4 \cdot T_8  \int dU e^{-\phi} g_{SS}^2 g_{xx}^{\frac{3}{2}}
    \sqrt{g_{00}g_{UU} - \left(2\pi \a' \hat{A_0}'\right)^2} \ ,
\end{eqnarray}
where $N_s = 2$ and reflects the two contributions from D8
and $\DeB$ branes and $\Omega_4$ results from the trivial angle
integration of $S^4$. The symbol $'$ denotes the differentiation
with respect to $U$. Note that $g_{00}g_{UU}$ is the only place
where the confined and deconfined phases are distinguished.

In terms of the dimensionless variables defined as~\cite{Bergman0}
\begin{equation}
u=\frac{U}{R}\, , \quad x_4=\frac{X_4}{R} \, , \quad
\tau=\frac{X_0^E}{R} \, , \quad \hat a = \frac{2\pi \alpha' \hat
A}{\sqrt{2N_f}R} \ ,
\end{equation}
the DBI action reads
\begin{eqnarray}
  S_{\mathrm{DBI}} =
  \caln \int du u^4 \sqrt{ f(u)(x_4'(u))^2 + \frac{1}{u^3}
  \left( \left\{\begin{array}{c} \inv{f(U)} \\ 1 \end{array}\right\} - (\hat{a}_0'(u))^2 \right) }
  \quad  \begin{array}{l} \mathrm{confined}
   \\ \mathrm{deconfined} \end{array} \ , \nn
\end{eqnarray}
where
\begin{equation}
{\mathcal N}= \frac{N_s N_f T_8 R^5 \Omega_4 (\beta V_3) }{g_s} \
.
\end{equation}
$\beta$ is the inverse temperature, $V_3$ is the volume of
$\vec{X}$ space.

For simplicity in this paper we only consider the antipodal
configuration, in which the D8 branes and $\DeB$ branes are
maximally separated at the boundary. This implies that, in the
confined phase, the D8-$\DeB$ branes are connected at $U=\Ukk$,
and, in the deconfined phase, D8 and $\DeB$ branes are parallel to
each other extending to the black hole horizon. Since it also
implies that $x_4$ is constant ($x_4' = 0$) the DBI action is
reduced to
\begin{eqnarray}
  S_{\mathrm{DBI}} =
  \caln \int du u^{\frac{5}{2}} \sqrt{
   \left\{\begin{array}{c} \inv{f(U)} \\ 1 \end{array}\right\} - (\hat{a}_0'(u))^2  }
  \qquad  \begin{array}{l} \mathrm{confined}
   \\ \mathrm{deconfined} \end{array} \ ,
\end{eqnarray}
The antipodal configuration simplifies the problem since we only
need to solve the one variable ($\hat{a}_0$) equation instead of
the coupled equations of two variables ($\hat{a}_0$, $x_4$). It
also simplifies the phase diagram of holographic QCD since there
is no ``cusp'' configuration studied in~\cite{Bergman0}. Thus the
deconfined phase at $T>T_c$ and any $\mu$ corresponds to the hot
QGP phase,~\footnote{In general there could also be other phases corresponding to connected configurations in
deconfined phase as shown in~\cite{Bergman0}.} and the confined phase at $T<T_c$ and sufficiently
large $\mu>\mu_c$ corresponds to a cold dense phase made of baryonic matter while at small $\mu<\mu_c$ the phase
is the trivial vacuum configuration with zero baryonic
density. The onset chemical potential, $\mu_c$, is defined in (\ref{muqgp1}). The antipodal
configuration will allow us to obtain analytic results for both
the hot QGP phase and the cold dense phase, which shall remain
qualitatively the same for large non-maximal separations of $\De$-$\DeB$
branes.

\subsection{Grand potential}

The grand potential is identified with the on-shell DBI action.
For convenience we will compute the rescaled grand potential
defined as~\footnote{The grand potential in SS model has been computed in several papers~\cite{FiniteD,Bergman0,Kim:Dense} with various notational conventions. In this paper we follow the convention in~\cite{Bergman0}.}
\begin{eqnarray}
  \Omega(t,\mu) = \inv{\caln}S_{\mathrm{DBI}}|_\mathrm{on-shell} \ ,
\end{eqnarray}
where $t$ is the dimensionless temperature and $\mu$ is the dimensionless chemical potential defined as%
\begin{eqnarray}
  t= T\cdot R\ , \qquad \mu = \hat{a}_0(\infty) \ .
\end{eqnarray}

\subsubsection{QGP phase}

Let us first consider the deconfined phase,
\begin{eqnarray}
  S_{\mathrm{DBI}} =
  \caln \int du u^{\frac{5}{2}} \sqrt{ 1 - (\hat{a}_0'(u))^2  }  \ .
\end{eqnarray}
Since $\hat{a}_0$ is a cyclic coordinate, its conjugate momentum
is conserved, which is defined up to a normalization as
\begin{eqnarray}
  d \equiv -\inv{\caln}\frac{\d S_\mathrm{DBI}}{\d \hat{a}_0'(u)}
  = u^{\frac{5}{2}} \frac{\hat{a}_0'}{\sqrt{
  1 - (\hat{a}_0'(u))^2  }   } \ ,
\end{eqnarray}
and $\hat{a}_0$ is easily solved. Especially
\begin{eqnarray}
  \mu = a_0(\infty) = \int_{u_T}^\infty d {u} \frac{d}
  {\sqrt{{u}^5 + d^2}} \ , \label{muqgp}
\end{eqnarray}
where $a(u_T) = 0$ for the regularity at the horizon. With this
solution the grand potential reads
\begin{eqnarray} \label{eqn_omega_QGP}
  \Omega(t,\mu) = \int_{u_T}^\infty du
  \frac{u^5}{\sqrt{u^5+d(\mu)^2}} \ ,
\end{eqnarray}
where $d(\mu)$ is the function of $\mu$ via (\ref{muqgp}).

\subsubsection{Cold dense phase}

In the confined phase we can do the same analysis with
\begin{eqnarray}
  S_{\mathrm{DBI}} =
  \caln \int du u^{\frac{5}{2}} \sqrt{ f(u) - (\hat{a}_0'(u))^2  }  \ .
\end{eqnarray}
However in this case we need to consider the explicit source term.
Contrary to the deconfined phase, D8 and $\DeB$ branes are connected at $u=u_{KK}$. For a nontrivial $\hat{a}_0$ there must be a
singularity at $u=u_{KK}$ and to take this account we consider the D4 branes wrapping $S^4$ as the baryon source. The source of a uniform distribution of D4 branes ($\sim d$) can be explicitly introduced by  the CS action. Referring to~\cite{Bergman0} for more detail,
we simply quote the final result for the grand potential and chemical potential %
\begin{eqnarray} \label{eqn_omega_Dense}
  \Omega(\mu) = \int_{u_{KK}}^\infty du \inv{\sqrt{f(u)}}
  \frac{u^5}{\sqrt{u^5+d(\mu)^2}} \ ,
\end{eqnarray}
\begin{eqnarray}
  \mu = a_0(\infty) = \int_{u_{KK}}^\infty d {u} \inv{\sqrt{f(u)}}
  \frac{d}{\sqrt{{u}^5 + d^2}} + \mu_c \ .\label{muqgp1}
\end{eqnarray}
where $\mu_c\equiv \frac{1}{3}u_{KK}$ is the onset chemical
potential of the cold dense phase.

\section{Baryonic Density and Susceptibilities in the QGP Phase}

At $T>T_c$ the Sakai-Sugimoto model has a deconfined, chirally
symmetric QGP phase for any $\mu$. We make use of the analytic
results for QGP phase to analyze its baryonic degrees of freedom.
Using Eq.(\ref{eqn_omega_QGP},\ref{muqgp}) and subtracting out the
vacuum part in the grand potential, we explicitly write out the
following equations as our starting point at given $T,\mu$:
\begin{equation}
P_{QGP}\left[T,d(T,\mu)\right]= \left[ \frac{2}{7}\Gamma_A
d^{\frac{7}{5}} + \frac{2}{7}u_T (d^2+u_T^5)^{\frac{1}{2}} -
\frac{2}{7}u_T d \,\, _2{\mathbf F}_1
\left(\frac{1}{5},\frac{1}{2};\frac{6}{5};-\frac{u_T^5}{d^2}\right)
\right]
\end{equation}
\begin{equation} \label{eqn_QGP_cons}
\Gamma_A d^{\frac{2}{5}}-u_T \,\, _2{\mathbf F}_1
\left(\frac{1}{5},\frac{1}{2};\frac{6}{5};-\frac{u_T^5}{d^2}\right)
-\mu=0  \, , \quad
\Gamma_A=\frac{\Gamma(\frac{3}{10})\Gamma(\frac{6}{5})}{\sqrt{\pi}}
\end{equation}
In principle we need to solve $d(T,\mu)$ from the constraint and
then obtain the full $P[T,\mu]$ to calculate density and
susceptibilities. However, we can make use of the chain rule to do
the calculation without solving the constraint. In other words, we
(temporarily) consider $\mu$ as being determined by $T,d$:
\begin{equation}
\mu=\Gamma_A d^{\frac{2}{5}}-u_T \,\, _2{\mathbf F}_1
\left(\frac{1}{5},\frac{1}{2};\frac{6}{5};-\frac{u_T^5}{d^2}\right)
\end{equation}
And we first evaluate the derivatives over $d$:
\begin{equation}
\frac{\partial P}{\partial d} = \frac{2 u_T}{3} \left(d^2\over
u_T^5\right)^{\frac{1}{2}} \,\, _2{\mathbf F}_1 \left( {3\over
10},{3\over 2};{13\over 10};-{d^2\over u_T^5} \right)
\end{equation}
\begin{equation}
\frac{\partial \mu}{\partial d} = \frac{2 u_T}{3d} \left(d^2\over
u_T^5\right)^{\frac{1}{2}} \,\, _2{\mathbf F}_1 \left( {3\over
10},{3\over 2};{13\over 10};-{d^2\over u_T^5} \right)
\end{equation}
The density is then given by $n_B={\partial P /\partial d \over
\partial \mu /\partial d}=d $, as it should. To determine its
ultimate dependence on $T,\mu$ we will have to solve the
constraint.

We now calculate the susceptibilities $d_n$ as defined in
Eq.(\ref{eqn_susceptibilities_def}). To do that we introduce
$P_n(T,d) \equiv {\partial P^n \over \partial \mu^n}$ with
$d_n=t^{n-4}\cdot P_n(d \to 0)$ (noting that $d\to 0$ is
equivalent to $\mu\to 0$). $P_n$ can be evaluated order by order,
using $P_{n+1}=\frac{\partial P_n / \partial d}{\partial \mu /
\partial d}$. The final results to the order $n=10$ are given below:
\begin{eqnarray}
d_{n=2,4,6,8,10} = \chi_n \,  u_T^{-n+7/2} \, t^{n-4} && = \chi_n
\, \left({4\pi\over 3}\right)^{-2n+7} \, t^{-n+3}
  \nonumber \\
\chi_2={3\over 2} \quad , \quad \chi_4=\frac{3^5}{2^3\cdot 13}
\quad , && \quad \chi_6= - \frac{3^8 \cdot 5 \cdot 31}{2^5\cdot
13^2\cdot 23}
\quad , \nonumber \\
  \chi_8= \frac{3^9\cdot 5^2\cdot 7\cdot 3011}{2^7\cdot 11\cdot 13^3\cdot
  23} \quad , && \quad \chi_{10} = -\frac{3^14\cdot 5^2\cdot 7\cdot 24546787}{2^9\cdot 11\cdot 13^4\cdot 23^2\cdot 43}
\end{eqnarray}
These results show very distinctive patterns: (\textbf{a}) first
of all we see simple power dependence on temperature but different
orders depend on $T$ very differently; (\textbf{b}) particularly,
only $d_2$ has positive power of $T$-dependence, i.e. growing with
$T$, while all higher order susceptibilities have negative power
thus vanish in the $T\to \infty$ limit; (\textbf{c}) furthermore,
we notice there is an alternating sign pattern for $n>2$, i.e.
$d_{4,8}$ are positive while $d_{6,10}$ are negative; (\textbf{d})
finally we notice the ratios between successive susceptibilities
are $d_{n+2}/d_n \sim t^{-2}$. As is evident from the above
results, even in the QGP phase with quarks as the basic baryonic
charge carriers, the strong interaction modifies the behavior
significantly, resulting in non-perturbative patterns of these
susceptibilities.

At this point, it would be interesting to see the actual
dependence of the susceptibilities on physical parameters by
recovering all the dimensions (and neglected constants) of
involved quantities, i.e. $P,\mu,T$. This can be done via the
following:
\begin{equation} \label{eqn_recover}
P\to P\times \left[ \frac{N_s\cdot N_f\cdot N_c\cdot
\lambda^3}{2^8\cdot 3\cdot \pi^5 \cdot M_{KK}^3\cdot R^7 } \right]
\quad , \quad \mu \to \mu \times \left[ \frac{\lambda}{4\cdot
\pi\cdot M_{KK}\cdot R^2} \right]
\end{equation}
We also re-scale temperature by $t\equiv TR \to \frac{T}{T_c}
\times [\frac{M_{KK} \cdot R}{2\pi}]\equiv \tilde{T} \times
[\frac{M_{KK} \cdot R}{2\pi}]$. Eventually we arrive at the
results below: \footnote{While we only worked out the
susceptibilities explicitly to the 10th order, it is plausible
that the dependence on $\lambda$ and $\tilde{T}$ is valid also for
even higher orders.}
\begin{equation} \label{eqn_QGP_results}
d_n = \xi_n \times N_s \times N_c \times N_f \times \left(1\over
\lambda \tilde{T} \right)^{n-3}
\end{equation}
with $\xi_n=\chi_n \cdot 3^{2n-8}/ (\pi\cdot 2^{n-3})$ (note the
sign patterns of $\xi_n$ follow from $\chi_n$, i.e.
$\xi_{2,4,8,...}>0$ while $\xi_{6,10,...}<0$). First of all we
notice the degrees of freedom counting reflects quark-like
dependence, as baryonic or any other non-single combinations of
quarks would result in different dependence on both $N_c$ {\em
and} $N_f$. On the other hand we see nontrivial power dependence
on the coupling: again $d_2 \sim \lambda \tilde{T}$ grows both
with $\lambda$ and $\tilde{T}$ but all higher susceptibilities
will be {\em suppressed at very strong coupling}.

Finally let's discuss the asymptotic behavior of baryonic density
at very large and very small density/chemical potential, i.e. $\mu
\to \infty$ and $\mu \to0$. In these limits we can solve the
constraint
equation to obtain the density.\\
\textbf{(i) Dense Limit in QGP Phase:} In the dense limit $\mu \to
\infty$ and thus $d\to \infty$, we solve the constraint equation
Eq.(\ref{eqn_QGP_cons}) to leading order and obtain the pressure
and baryonic density to be (after recovering physical dimensions):
\begin{equation}
\frac{P^{dense}}{T_c^4}\sim N_s \cdot N_f \cdot N_c
 \cdot \lambda^{-{1\over 2}} \left(\frac{\mu}{T_c}\right)^{7\over 2} \,\, ,
 \quad  \frac{n_B^{dense}}{T_c^3} \sim \frac{7}{2}\cdot N_s \cdot N_f \cdot N_c
 \cdot \lambda^{-{1\over 2}} \left(\frac{\mu}{T_c}\right)^{5\over 2}
\end{equation}
We can also calculate the energy per particle
\begin{equation}
\bar E \approx \frac{\mu n -P}{n} = \frac{5}{7} \, \mu =
\frac{20}{21}\, \bar E_{S.B.}
\end{equation}
with $E_{S.B.}=3\mu/4$ representing the energy per particle for a
free gas of massless fermions at zero temperature and finite
$\mu$. Amusingly, the strong coupling result misses the
non-interacting limit by only $1/21 \sim 5\%$ despite the strong coupling dynamics. \\
\textbf{(ii) Dilute Limit in QGP Phase:} In the dilute limit $\mu
\to 0$ and thus $d \to 0$, we note that $\left[\, _2{\mathbf F}_1
\left(\frac{1}{5},\frac{1}{2};\frac{6}{5};-\frac{u_T^5}{d^2}\right)
\right] {\bigg |}_{d\to 0} \to \Gamma_A \cdot
\frac{d^{\frac{2}{5}}}{u_T} -
\frac{2}{3}\frac{d}{u_T^{\frac{5}{2}}} + \hat o(d^2)$ and can then
calculate the baryonic density to leading order
\begin{equation}
\frac{n_B^{dilute}}{T_c^3} \sim  N_s N_f N_c \lambda
\left(\frac{T}{T_c}\right)^3 \left(\frac{\mu}{T_c}\right)
\end{equation}
This result has linear dependence on $\mu$, in common with the
leading order at small $\mu$  of free fermion gas at UR limit. The
strong coupling density here however depends on coupling $\lambda$
and also has $T^3$ dependence, differing from the free case which
has no coupling and has $T^2$ dependence.

\section{Baryonic Density and Susceptibilities in the Cold Dense Phase}

We now turn to the cold dense phase at $T=0$. To study this phase,
we focus on the situation with maximal $D8$-$\bar D 8$ separation
which allows tractable analytic formulae and much simplifies the
calculation. For non-maximal separation, the physics at $T=0$
remains qualitatively the same while the calculation is much more
involved.

We start with Eq.(\ref{eqn_omega_Dense},\ref{muqgp1}) and further
rewrite the pressure and chemical potential as a function of $d$
into the following forms by re-scaling the quantities with
$u_{KK}$:
\begin{eqnarray} \label{eqn_scaled_P}
\tilde{P}=\frac{P}{\left(u_{KK}\right)^{\frac{7}{2}}}=\int_1^\infty
d\tilde{u} \frac{\tilde{u}^4}{\sqrt{\tilde{u}^3-1}}
\left[\frac{1}{\sqrt{1+\frac{\tilde{d}^2}{\tilde{u}^5}}}-1 \right]
\end{eqnarray}
\begin{eqnarray} \label{eqn_scaled_mu}
\tilde{\mu}= \frac{\mu}{u_{KK}} = \tilde{\mu_c} + \int_1^\infty
d\tilde{u} \frac{\tilde{u}^{\frac{3}{2}}}{\sqrt{\tilde{u}^3-1}}
\sqrt{\frac{\tilde{d}^2}{\tilde{u}^5+\tilde{d}^2}}
\end{eqnarray}
In the above $\tilde{d}^2\equiv d^2/u_{KK}^5$ and
$\tilde{\mu_c}=\mu_c/u_{KK}=1/3$. Also note in the pressure we
have subtracted out the vacuum part, e.g. the $(-1)$ within the
bracket. These can be easily solved numerically: for each given
$\tilde{\mu}$ we can first solve $\tilde{d}$ from
Eq.(\ref{eqn_scaled_mu}) and then obtain $\tilde{P}$ from
Eq.(\ref{eqn_scaled_P}), and the results are shown as solid blue
lines in Fig.2.

A few comments are in order. First we notice that for
$\tilde{\mu}<\tilde{\mu_c}$ there is no solution for $\tilde{d}$
as is evident from Eq.(\ref{eqn_scaled_mu}), so the solution is
the trivial U-shape and the system is in vacuum phase without any
dependence on $\tilde{\mu}$. When $\tilde{\mu} > \tilde{\mu_c}$
baryons start to emerge and a new phase with nonzero baryonic
density takes over the vacuum one. Furthermore both the pressure
and the baryonic density are continuous at $\tilde{\mu_c}$
transition while the first derivative of $\tilde{d}$ versus $\mu$
(i.e. the lowest susceptibility at $\tilde{\mu_c}$) is not, which
implies a second order phase transition. In physical unit, one has
$\mu_c=\frac{\lambda M_{KK}}{27\pi}=\frac{2}{27}\times \lambda
\times T_c$: with $\lambda\sim 20-50$, the relation reasonably
agrees with current rough estimate of $\mu_c/T_c$ in QCD.

We now make an expansion of the pressure and density in the cold
dense phase close to $\tilde{\mu_c}$. This can be done by
systematically analyze the expansion of $\tilde{d}\to 0$ order by
order in the integrands of
Eqs.(\ref{eqn_scaled_P},\ref{eqn_scaled_mu}). The result for the
pressure is:
\begin{eqnarray}
\tilde{P}&&= \sum_n c_{2n}
\left(\tilde{\mu}-\tilde{\mu_c}\right)^{2n}
 \\
c_2 && =\frac{3}{2\pi} \, , \,\, c_4 = \frac{9\,
\Gamma\left(\frac{13}{6}\right)}{\pi^{\frac{7}{2}}\,
\Gamma\left(\frac{11}{3}\right)} \, , \,\, c_6=\frac{243}{16\pi^6}
\left[ \frac{2\,
\Gamma\left(\frac{13}{6}\right)^2}{\Gamma\left(\frac{8}{3}\right)^2}
- \frac{\sqrt{\pi}\,
\Gamma\left(\frac{23}{6}\right)}{\Gamma\left(\frac{13}{3}\right)}
\right] \, , \,\, ... \nonumber
\end{eqnarray}
We skip to show the higher order coefficients whose expressions
become really long. Here we give an idea of the numbers:
$c_2=0.477465,c_4=0.0441772,c_6=0.001576,c_8=-0.000365984,c_{10}=0.0000107106$.
These coefficients can be directly related to baryonic
susceptibilities defined at $\tilde{\mu}=\tilde{\mu_c}$, i.e.:
\begin{equation}
D_{2n}\equiv \frac{\partial^{2n} (P/\mu_c^4)}{\partial (\mu /
\mu_c)^{2n}}{\bigg |} _{\mu=\mu_c} = \frac{\tilde{\mu_c}^{2n-4}
}{u_{KK}^{\frac{1}{2}}} \frac{\partial^{2n} \tilde{P}}{\partial
\tilde{\mu}^{2n}} {\bigg |} _{\mu=\mu_c}
=\frac{\tilde{\mu_c}^{2n-4} }{u_{KK}^{\frac{1}{2}}} \, (2n)! \,
c_{2n}
\end{equation}
In Fig.2(left) we compare the full numerical result (solid blue
line) of the pressure with its Taylor series results truncated at
the 1st(red), 2nd(green), 3rd(orange), 4th(magenta), and
5th(purple) order respectively. Interestingly we found that the
expansion till the 2nd order, i.e. the series with
$(\tilde{\mu}-\tilde{\mu_c})^2$ and
$(\tilde{\mu}-\tilde{\mu_c})^4$ terms only, agrees remarkably well
with the full result up to very large $\tilde{\mu}$, indicating
some fine cancellations among all higher order terms.

\begin{figure*}
    \hskip 0in
\center{    \includegraphics[width=6.6cm]{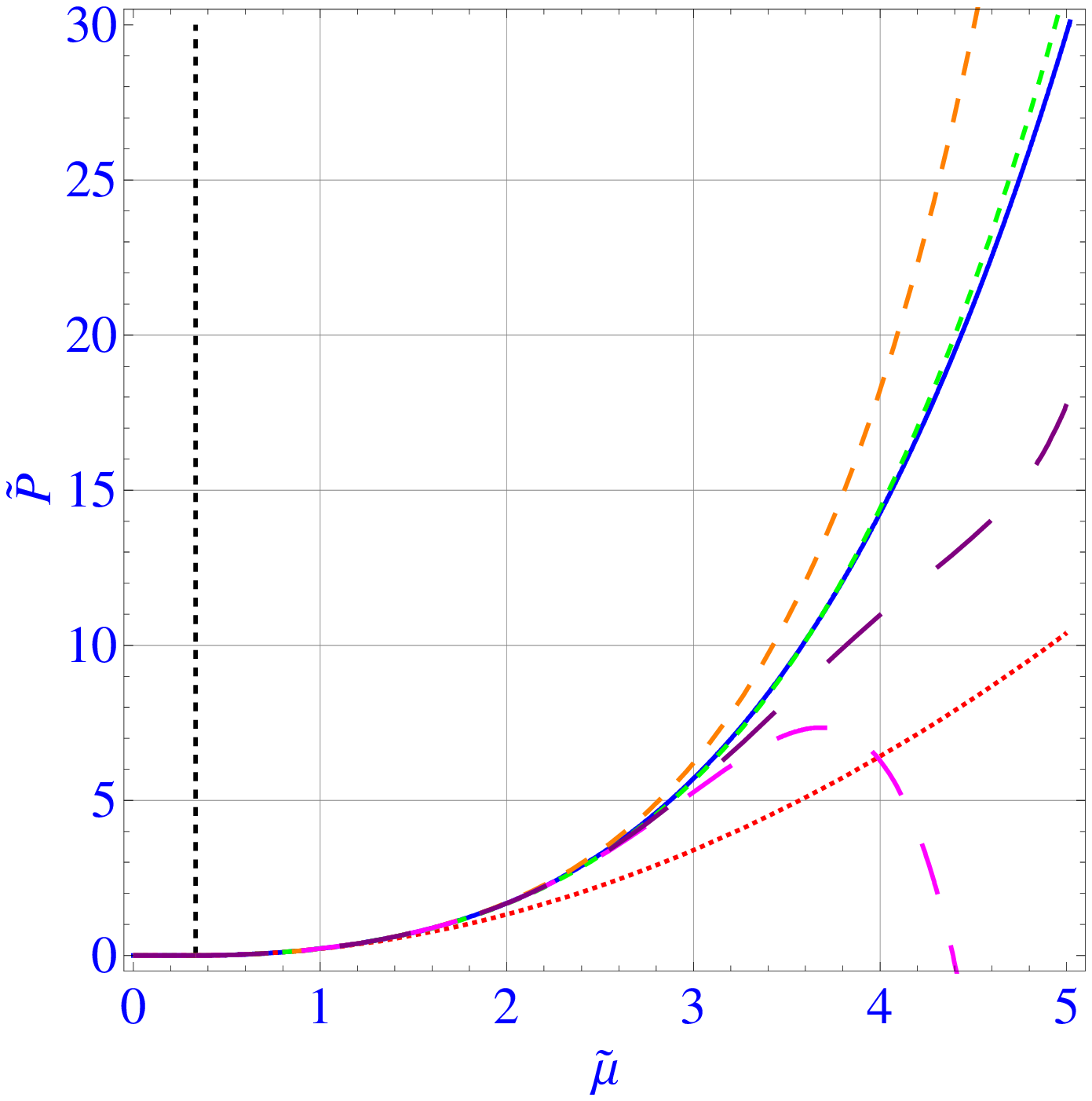}
    \hskip 0.1in
   \includegraphics[width=6.6cm]{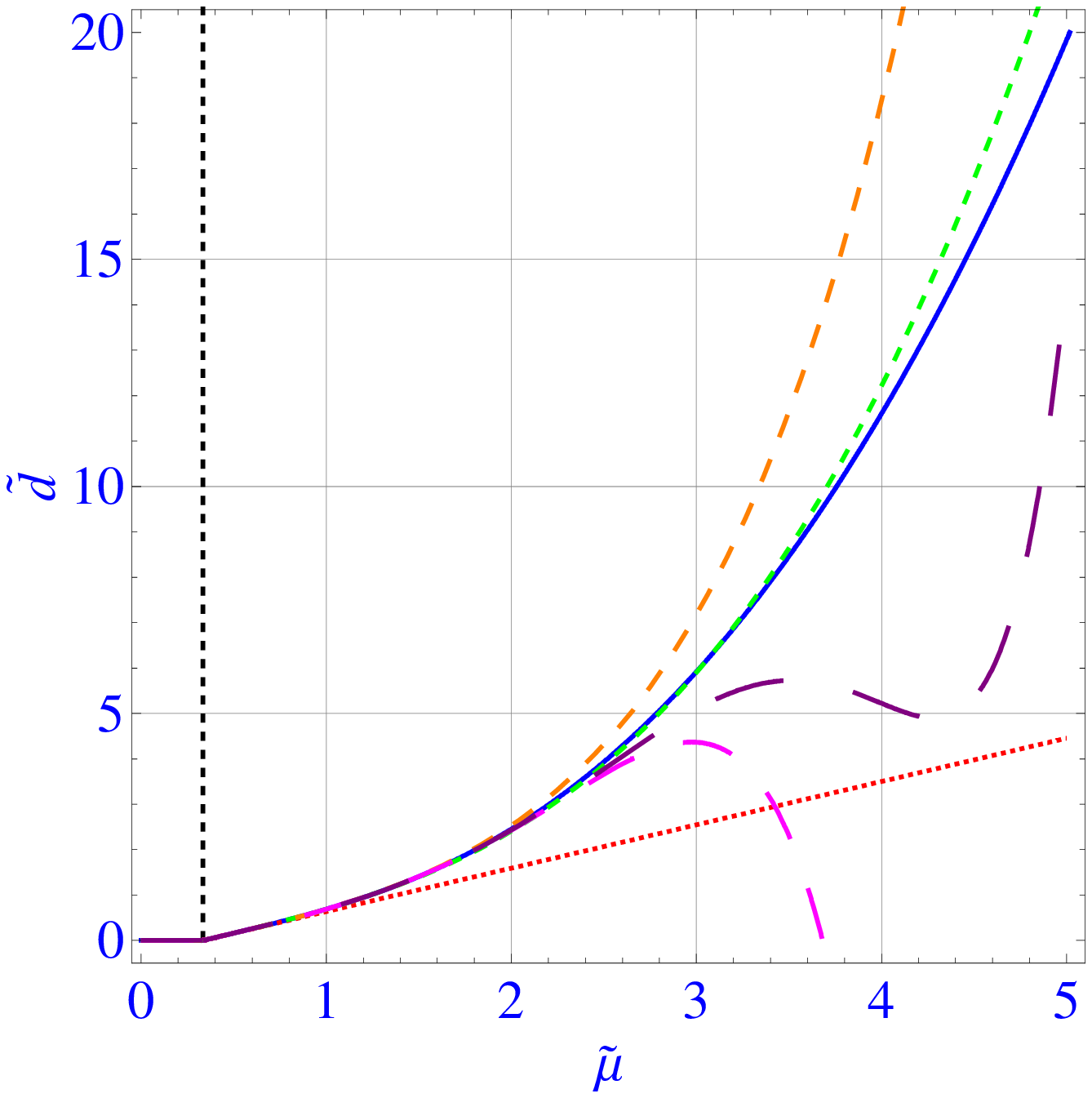} }
 \caption{ Comparison of the full numerical results(solid blue line)
 for the pressure(left) and density(right) with their respective Taylor series
 results(dashed lines with growing lengths) truncated at the 1st(red), 2nd(green), 3rd(orange), 4th(magenta),
 and 5th(purple) order. The vertical dashed black line indicates the position of the phase transition.}
 \end{figure*}

\begin{figure}
    \hskip 0in
\center{    \includegraphics[width=7.5cm]{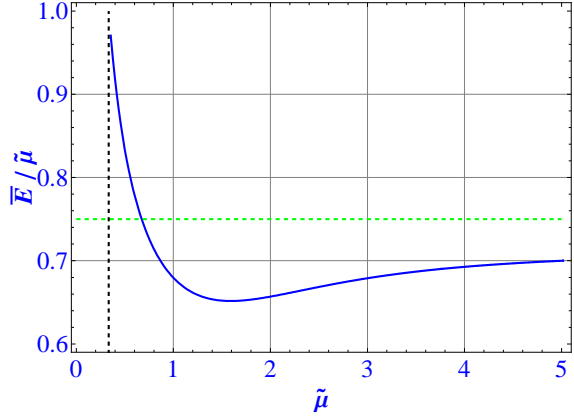}}
 \caption{The energy per particle versus chemical potential. The horizontal
 dashed green line indicates the free value. The vertical dashed black line
 indicates the position of the phase transition. }
 \end{figure}

The expansion for the density is given by:
\begin{eqnarray}
\tilde{d}&&= \sum_n f_{2n-1}
\left(\tilde{\mu}-\tilde{\mu_c}\right)^{2n-1}
 \\
f_1 && =\frac{3}{\pi} \, , \,\, f_3 = \frac{36\,
\Gamma\left(\frac{13}{6}\right)}{\pi^{\frac{7}{2}}\,
\Gamma\left(\frac{11}{3}\right)} \, , \,\, f_5=\frac{729}{8\pi^6}
\left[ \frac{2\,
\Gamma\left(\frac{13}{6}\right)^2}{\Gamma\left(\frac{8}{3}\right)^2}
- \frac{\sqrt{\pi}\,
\Gamma\left(\frac{23}{6}\right)}{\Gamma\left(\frac{13}{3}\right)}
\right] \, , \,\, ... \nonumber
\end{eqnarray}
Again we skip to show the higher order coefficients and give an
idea of the numbers:
$f_1=0.95493,f_3=0.176709,f_5=0.00945598,f_7=-0.00292787,f_9=0.000107106$.
From the expansion it is clear that there is a jump in the slope
of $\tilde{d}$ versus $\mu$, i.e. $0$ at $\tilde{\mu}\to
\tilde{\mu_c}^-$ and $\frac{3}{\pi}$ at $\tilde{\mu}\to
\tilde{\mu_c}^+$. We also note that these coefficients are related
to those in the pressure expansion by $f_{2n-1}=(2n)\cdot c_{2n}$
order by order as required by $\tilde{d}=\partial_{\tilde{\mu}}
\tilde{P}$. In Fig.2(right) we compare the full numerical result
of the density (solid blue line) with its Taylor series results
truncated at the 1st(red), 2nd(green), 3rd(orange), 4th(magenta),
and 5th(purple) order respectively. Not surprisingly we again
found that the expansion till the 2nd order, i.e. the series with
$(\tilde{\mu}-\tilde{\mu_c})$ and $(\tilde{\mu}-\tilde{\mu_c})^3$
terms only, agrees remarkably well with the full result up to very
large $\tilde{\mu}$.

Finally we study another quantity of interest, i.e. the energy per
particle $\bar{E}$ which at $T=0$ is given by $\bar{E} =
(\tilde{\mu} \tilde{d} - \tilde{P})/\tilde{d} \equiv
g(\tilde{\mu}) \tilde{\mu}$. The function $g(\tilde{\mu})$ is
plotted in Fig.3 as solid blue line. The dashed green line
indicates the free case $g=3/4$. We see that close to the
transition point the energy per particle deviates much from the
free value and drops very sharply, while at much larger density it
curves back and approaches the free value from below very slowly.
We notice that the curve crosses the free line at about
$\tilde{\mu}\approx 2\tilde{\mu_c}$: it may imply that below this
density the system has repulsive interaction and above it the
system has attractive interaction. In either regions, the
deviation of $\bar{E}$ from free case is rather modest.

We also present the results with physical scales recovered as in
Eq.(\ref{eqn_recover})
\begin{eqnarray}
P &=& \frac{1}{2\cdot 3^8\cdot \pi^5} \times N_s\times N_f\times
N_c \times (\lambda^3 M_{KK}^4) \times \sum_n
\frac{c_{2n}}{3^{2n}}
\left(\frac{\mu}{\mu_c}-1 \right)^{2n} \nonumber \\
&=& N_s\times N_f\times N_c \times (\frac{\mu_c^4}{2\pi \lambda})
\times \sum_n \frac{c_{2n}}{3^{2n-4}} \left(\frac{\mu}{\mu_c}-1
\right)^{2n}
\end{eqnarray}
and
\begin{eqnarray}  \label{eqn_dense_results}
D_{2n}\equiv \frac{\partial^{2n} (P/\mu_c^4)}{\partial (\mu /
\mu_c)^{2n}}{\bigg |} _{\mu=\mu_c} = N_s\times N_f\times N_c
\times \frac{c_{2n} (2n)!}{3^{2n-4} (2\pi)} \times
\frac{1}{\lambda}
\end{eqnarray}
where we have used $\mu_c=\frac{u_{KK}}{3}=\frac{\lambda
M_{KK}}{27 \pi}$.

\section{Summary and Discussions}

In summary, we have calculated the baryonic density and
susceptibilities in a holographic model of QCD. The results both
for the hot QGP phase in Eq.(\ref{eqn_QGP_results}) and for the
cold dense phase in Eq.(\ref{eqn_dense_results}) show interesting
patterns due to strong coupling dynamics. In the following we
discuss relevant lessons  for QCD that can be learned from the
results.

We first discuss the results for hot QGP phase close to $T_c$. To
give an idea of the numbers, we re-write the results in
Eq.(\ref{eqn_QGP_results}) as:
\begin{equation}
\frac{d_2}{N_s N_f N_c}\approx 0.012 \cdot \lambda \tilde{T} \,\,
, \quad \frac{d_4}{N_s N_f N_c}\approx \frac{0.37}{\lambda
\tilde{T}}
 \,\, , \quad \frac{d_6}{N_s N_f N_c}\approx -\frac{26}{\lambda^3
\tilde{T}^{\, 3}}
\end{equation}
We first note that a few qualitative features agree well with the
lattice QCD data shown in Fig.\ref{fig_lattice}: the sign pattern
of $d_{2,4,6}$ is the same, $d_2$ approximately exhibits linear
growth with $T$ close to $T_c$ while $d_{4,6}$ shows quick
decrease with $T$, and also $d_6$ vanishes much more abruptly than
$d_4$. If one takes a $\lambda\sim 10-20,$\footnote{$\lambda\sim
17$ from \cite{Sakai}.} our $d_2$ values are very close to the
lattice results, but the $d_4$ and $d_6$ values are too small. The
comparison seems to indicate that dynamics from very strong
coupling of {\em quarks only} tends to suppress higher order
susceptibilities and may not be adequate to account for the
$d_{4,6}$ close to $T_c$ seen in lattice QCD \footnote{As a
caveat, one should be aware that already around the $T_c$ scale,
the many (un-included) higher KK modes may start kicking in and
contributing to the susceptibilities --- these become important
with even higher temperature. At present it is not clear whether
including their contributions would improve the agreement with
lattice QCD results.} : this is yet another indication that those
higher order susceptibilities may be attributed to multi-quark
correlations like baryons which manifest themselves more and more
in higher order susceptibilities as we already pointed out in
Section.II. An even more critical quantity to further distinguish
the two types of contribution to susceptibilities may be the
non-diagonal susceptibilities \cite{Koch:2008ia} after introducing
isospin chemical potential for quark flavors. These can be
implemented in the holographic model of QCD
\cite{Parnachev:2007bc}, and will be studied elsewhere. These
susceptibilities can also be suitably combined to give the
baryonic charge fluctuations in the QGP, which have
phenomenological importance and can be measured in RHIC
experiments \cite{Koch:2008ia}: our results imply that strong
coupling of quarks may enhance event-by-event fluctuation of
baryon number as seen in the covariance $<\delta B^2>\sim \lambda
\tilde{T}$ while suppress even higher order variances.

Another interesting observation is that for both phases we have
noticed that in the dense regime the energy per particle $\bar E$
is very close to the Stefan-Boltzmann limit, i.e. $\frac{\bar
E}{\bar E_{SB}}$ (the so-called Bertsch number) is close to 1
despite the strong coupling. We remind that the result of entropy
in $N_c$ D3 branes system (roughly involving adjoint gauge sector
but without fundamental fermions) deviates from the
Stefan-Boltzmann limit by $25\%$. This comparison indicates that
the strong coupling dynamics may somehow modify the thermodynamics
of fundamental fermions much more mildly than of the gauge fields.
Interestingly lattice QCD seems to show similar situation: while
$d_2$ for fermionic thermodynamics approaches the Stefan-Boltzmann
limit already around $1.4T_c$, the pure gauge lattice results for
gluonic thermodynamics show no tendency toward the same limit even
at $4T_c$.

Finally we point out that the cold dense phase of the
Sakai-Sugimoto model studied above has its many features similar
to the quarkyonic matter in large $N_c$ QCD
\cite{McLerran:2007qj}\cite{McLerran:2008ua}. Both are in the
confined regime but have their thermodynamics scaling as $N_c N_f$
like a quark system. The transition from the vacuum phase to the
dense phase for both happens at about the baryon mass threshold
\footnote{The transition order however is the 1st in QCD while the
2nd in the present holographic model: the difference may be due to
the absence of nucleon-nucleon interaction in the latter approach
where the baryons are only introduced as static sources. See also
comments in \cite{Bergman0}. }, after which the baryonic density
starts growing. While calculation is hard for the quarkyonic
matter in QCD, one may study its holographic dual quantitatively
and obtain useful results as ours in Section.V. It will be very
interesting to further investigate via the holographic method the
many interesting features of the quarkyonic matter, e.g. the
chiral symmetry and excitations near or deep into the Fermi
surface, etc, which were previously qualitatively studied.

\vspace{0.25in}

{\em \textbf{Acknowledgements:}} We thank Ismail Zahed, Sang-Jin
Sin, Andrei Parnachev, and Matthew Lippert for discussions. J.L.
is also grateful to Volker Koch, Larry McLerran, Edward Shuryak
and Frithjof Karsch for helpful comments. J.L. is supported by the
Director, Office of Energy Research, Office of High Energy and
Nuclear Physics, Divisions of Nuclear Physics, of the U.S.
Department of Energy under Contract No. DE-AC02-05CH11231. K.K. is
supported in part by US-DOE grants DE-FG02-88ER40388 and
DE-FG03-97ER4014.

\appendix

\section{The Background Metric Generated by D4 branes}
In this appendix we summarize the metric generated by $N_c$ D4 branes presented in \cite{Kruczenski}.

At zero temperature, the metric generated by $N_c$ D4-branes are given by
\begin{eqnarray}
&&ds^2=\left(\frac{U}{R}\right)^{3/2} \left(-(dX_0)^2  +
(d\vec{X})^2+f(U)(dX_4) ^2\right)  \nn \\
&& \qquad
+\left(\frac{R}{U}\right)^{3/2} \left(\frac{dU^2}{f(U)}+U^2 d\Omega_4^2\right)\  , \nn\label{LowT}
\end{eqnarray}
where
\begin{eqnarray}
  f(U) \equiv
1-\frac{\Ukk^3}{U^3} \ . \label{fkk}
\end{eqnarray}
$\vec{X}= X_{1,2,3}$, while $U (\geq U_{KK})$ and $\Omega_4$ are
the radial coordinate and four angle variables in the
$X_{5,6,7,8,9}$ direction. $R$ is given by
\begin{eqnarray}
  R^3 \equiv \pi g_s N_c l_s^3 \ , \label{R3}
\end{eqnarray}
where $g_s$ and $l_s$ are the string coupling and length
respectively. Since $X_4$ is compactified to a circle, to avoid a
conical singularity at $U=\Ukk$, the period of $\d X_4$ is set to
\begin{eqnarray}
\d X_4 = \frac{4\pi}{3}\frac{R^{3/2}}{\Ukk^{1/2}} \ .
\end{eqnarray}
%
%x
The field theory is defined by Kaluza-Klein mass ($\Mkk$) and the
four-dimensional coupling constant at the compactification scale,
$g_{YM}$,
\begin{eqnarray}
\Mkk \equiv
 \frac{2\pi}{\d X_4} = \frac{3}{2}
\frac{\Ukk^{1/2}}{R^{3/2}} \ , \qquad g_{YM}^{2} = \frac{g_5^2}{\d
X_4} = 2\pi \Mkk g_s l_s \ . \label{MkkGym}
%= 3 \sqrt{\pi} \left( \frac{g_{s}U_{KK}}{N_{c}l_{s}}\right)^{1/2}
\end{eqnarray}
where $g_5 (=  (2\pi)^2 g_s l_s) $ is the five-dimensional
coupling constant obtained from D4 brane DBI action. From
(\ref{R3}) and (\ref{MkkGym}) the parameters $R$, $\Ukk$, and
$g_s$ may be expressed in terms of $\Mkk$, $\l( = \gym^2 N_c)$,
and $l_s$ as
\begin{eqnarray}
R^3 =  \frac{\l  l_s^2}{2\Mkk}\ , \quad \Ukk = \frac{2}{9} \l \Mkk
l_s^2 \ , \quad g_s = \frac{\l}{2\pi\Mkk N_c l_s}\,\,. \
\end{eqnarray}

At finite-temperature, there are two possibilities
\cite{Witten:1998zw}. One is to follow the standard prescription
of the finite temperature field theory. The geometry is the same
as the zero temperature apart from the fact that the time
direction is Euclidean ($X_0 \ra X^{E}_0 = iX_0$) and compactified
with a circumference $\b = 1/T$:
\begin{eqnarray}
&&ds^2=\left(\frac{U}{R}\right)^{3/2} \left( (dX_0^{E})^2 +
(d\vec{X})^2 + f(U) (dX_4)^2\right) \nn \\
&& \qquad +\left(\frac{R}{U}\right)^{3/2} \left(\frac{dU^2}{f(U)}+U^2
d\Omega_4^2\right)\  , \nn
\end{eqnarray}
This corresponds to the confined phase which is thermodynamically
preferred (i.e. with the smallest action) in the low temperature
regime. The other possible geometry contains the black hole, which
is another saddle point of the Euclidean path integral over
supergravity configuration.  The pertinent background is
\begin{eqnarray}
&&ds^2=\left(\frac{U}{R}\right)^{3/2} \left(f(U)(dX^{E}_0)^2  +
(d\vec{X})^2+ (dX_4)^2\right) \nn \\
&& \qquad +\left(\frac{R}{U}\right)^{3/2}
\left(\frac{dU^2}{f(U)}+U^2 d\Omega_4^2\right)\ \nn ,
\label{HighT}
\end{eqnarray}
where\footnote{Strictly speaking we need to use the different
notation to distinguish (\ref{ft}) from (\ref{fkk}). However for
notational convenience we will use the same notation. It will not
make any confusion since the meaning will be clear from the
context. }
\begin{eqnarray}
  f(U)\equiv
1-\frac{U_T^3}{U^3} \ . \label{ft}
\end{eqnarray}
This corresponds to deconfined phase which is thermodynamically
preferred (i.e. with the smallest action) at high temperature. To
avoid a conical singularity at $U=U_T$ the period of $\d X^E_0$ of
the compactified $t_E$ direction is set to
\begin{eqnarray}
\d X^E_0 = \frac{4\pi}{3}\left( \frac{R^3}{U_T}\right)^{1/2}
\equiv \frac{1}{T} \equiv \b \ ,
\end{eqnarray}
which is identified with the inverse temperature. %
The confinement/deconfinement phase transition, i.e. the switching
from one background geometry to the other, occurs when $\d X_4 =
\d X^{E}_0$ \ie $\ $ at the critical temperature $T_{c}$,
\begin{eqnarray}
T_{c} = \frac{\Mkk}{2\pi} \ .
\end{eqnarray}

\end{document}